\documentclass{aa}
\usepackage{psfig}
\usepackage{natbib}
\newcommand\etal{{et al.~}}

\newcommand\aap{{A\&A}}

\newcommand\aj{{AJ}}
\newcommand\apj{{ApJ}}
\newcommand\apjl{{ApJ}}
\newcommand\apjs{{ApJS}}

\newcommand\pasj{{PASJ}}
\def\spose#1{\hbox to 0pt{#1\hss}}
\newcommand\simlt{\mathrel{\spose{\lower 3pt\hbox{$\mathchar"218$}}
     \raise 2.0pt\hbox{$\mathchar"13C$}}}
\newcommand\simgt{\mathrel{\spose{\lower 3pt\hbox{$\mathchar"218$}}
     \raise 2.0pt\hbox{$\mathchar"13E$}}}

\begin{document}

\title{Low mass T Tauri and young brown dwarf candidates in the Chamaeleon~II dark cloud found by DENIS}

\subtitle{}

\author{M.H. Vuong \inst{1}\thanks{Present address: Service d'Astrophysique, CEA Saclay, 91191 Gif-sur-Yvette, France}, L. Cambr\'esy \inst{2}, \and N. Epchtein \inst{1}}

\offprints{M\~y H\`a Vuong}

\institute{Observatoire de la C\^ote d'Azur, D\'epartement Fresnel, 06304 Nice Cedex, France; vuong@discovery.saclay.cea.fr, epchtein@obs-nice.fr \and California Institute of Technology, IPAC/JPL, CA 91109 Pasadena, USA; laurent@ipac.caltech.edu}
        
\date{Accepted 2001 September 5}
\authorrunning{Vuong et al.}
\titlerunning{Low mass T Tauri and young brown dwarf candidates}

\abstract{
We define a sample designed to select low-mass T Tauri stars and young brown dwarfs using DENIS data in the Chamaeleon~II molecular cloud. We use a star count method to construct an extinction map of the Chamaeleon~II cloud. We select our low-mass T Tauri star and young brown dwarf candidates by their strong infrared color excess in the $I-J/J-K_s$ color-color dereddened diagram. We retain only objects with colors $I-J \ge 2$, and spatially distributed in groups around the cloud cores . This provides a 
sample of 70 stars of which 4 are previously known T Tauri stars. We have carefully checked the reliability of all these 
objects by visual inspection on the DENIS images. Thanks to the association of the optical $I$ band to the infra-red $J$ and $K_s$ bands in DENIS, we can apply this selection method to all star formation regions observed in the Southern Hemisphere.
We also identify six DENIS sources with X-ray sources detected by {\it ROSAT}. 
Assuming that they are reliable low-mass candidates and using the evolutionary models for low-mass stars, we estimate the age of these sources between 1~Myr and $<$ 10~Myr.     
}
\maketitle

%
%________________________________________________________________

\section{Introduction}

The Chamaeleon complex is one of the nearest star formation regions to the Sun 
\citep[between 160 and 180 pc,][]{whi97}. It consists of three main dark clouds, designated Chamaeleon (Cha) I, II and III \citep[]{sch77}. Its relative proximity, high Galactic latitude (b $\approx -17 \degr$) and young age ($\sim 10^6$ yrs) make it an ideal location to search for low-mass T Tauri stars (TTS) and young brown dwarfs. The low-mass population can provide constraints on the shape of the initial mass function (by investigating the slope), which is still poorly determined at low masses \citep{tin93,mer96}.
%(Tinney 1993, M\'era et al. 1996, Gould et al. 1996)

Young stellar objects are characterised by their H$\alpha$ emission line \citep[]{har93} and an infrared excess that reveals the presence of circumstellar material in the near infrared \citep[]{lar98,oas99,gom01}, the mid-infrared \citep[ISO,][]{per00} and the far-infrared \citep[IRAS,][]{pru92}. X-ray observations (ASCA, ROSAT) allow the identification of young stellar objects \citep{yam98,alc00}. The origin of the X-ray emission from the low-mass pre-main sequence stars may be due to an enhanced dynamo activity with a coupled process of surface convection and differential rotation of a star or between a star and the inner disks \citep[{\it e.g.} in $\rho$ Oph,][]{mon00}. The millimeter observations such as the detection of the CO and CS lines that reveal the gas emission and the outflows often allow to characterise young stellar objects \citep[]{olm94}.

Weak line TTS are known as pre-main sequence objects without 
a disk and are usually found to be associated  with X-ray sources. In contrast, classical TTS with a circumstellar disk show strong 
H$\alpha$ emission, and excess IR emission.
Brown dwarfs are low-mass stars ($<$0.08 M$_{\sun}$) which never reach the reaction of hydrogen fusion. They are mainly powered by gravitational energy. Therefore they are relatively faint objects and can only be detected at nearby distances ($\simlt$ 30 pc). However, young brown dwarfs are much more luminous ($\sim$100 times more than field brown dwarfs). Such young brown dwarfs can be found in star forming regions out to a few 100 pc \citep{neu99}.

The Cha~II has not been mined as thoroughly as the Cha~I cloud.  
In the Chamaeleon clouds, the pre-main-sequence stars, mainly TTS, were first discovered by \citet{sch77}. No brown dwarfs have yet been found in the Cha~II. Recently, \citet{com00} found eleven young brown dwarfs in the Cha~I cloud from pointed IR observations.
With the advent of large near-IR surveys (DENIS, 2MASS), it is now possible to easily identify such samples across large regions of the sky.
\citet{cam98} have used DENIS data to find young stellar objects in the Cha~I cloud.

In this paper, we introduce a low-mass star sample in the Cha~II dark cloud, extracted from the DENIS survey. Sect.2 describes the observations and the data reduction. In Sect.3, we construct an extinction map of the Cha~II cloud, and present the color-color $I-J$/$J-K_s$ diagram used to search for new low-mass members. Finally, Sect.4 discusses the position of our sources on the color-magnitude diagram using the evolutionary tracks for low-mass stars as modeled by \citet{bar98}.

\section{DENIS Observations and data reduction}

The observations presented in this paper have been obtained from the DEep Near Infrared Survey of the Southern Sky \citep[DENIS,][]{epc97} between 1996 March and 1998 May at La Silla using the ESO 1m telescope. They cover an area of 2.42$\degr$  $\times$ 3.50$\degr$ centered at 13h 00m 00s in right ascension (J2000) and $-$76$\degr$ 45$\arcmin$ 00$\arcsec$ in declination (J2000) in three bands $I$(0.8$\mu m$), $J$(1.25$\mu m$) and $K_s$(2.15$\mu m$). Limiting magnitudes at 3$\sigma$ are 18, 16 and 13.5 in $I$, $J$ and $K_s$ bands.
It consists of 24 strips each containing 180 images of 12$\arcmin$ $\times$ 12$\arcmin$ taken at constant right ascension along an arc of 30$\degr$ in declination. There is a 2$\arcmin$ overlap between each two adjacent images. For the adjacent strips, the overlap is also 2$\arcmin$ for the image in the north. The overlap becomes more important when we move to the south by a simple projection effect (this is particularly important for the Chamaeleon complex because of its proximity to the South Pole).

The data reduction took place at the Paris Data Analysis Center (PDAC). The source extraction used PSF fitting. The astrometric calibration was obtained by cross-correlation with the USNO-PMM catalog \citep[]{mon98}. The absolute astrometry is then fixed by the accuracy of this catalog \citep[$\sim$ 0.5$\arcsec$ at 3$\sigma$,][]{deu99}. The internal accuracy of DENIS observations derived from the overlapping regions of adjacent images is $\sim$0.35$\arcsec$ (3$\sigma$). For the determination of the photometric zero points, all standard stars observed during a given night were used. We have checked the photometric uncertainties of sources in the Cha~II cloud by comparing the magnitudes of stars detected in more than one image. This yields photometric uncertainties of $\sim$ 0.05, 0.12 and 0.15 for $I$, $J$ and $K_s$, respectively.
All point sources detected in an image are taken into account in the PDAC procedures. Consequently, an object can appear several times, when it is located in the overlapping regions.
To eliminate these multiple detections, we first determine the radius in which two point sources are considered to be the same object. To determine this radius, we use the histogram of the position differences of sources in the overlapping regions between two adjacent images. The distribution drops to 0 identifications at a radius of 2$\arcsec$ before starting to pick up unrelated stars beyond 10$\arcsec$. Adopting a 2$\arcsec$ radius\footnote{Note that these objects are located at the edges of the images, and therefore, their positional accuracy is worse than those located in the centre.} assures that all double entries are removed, which is especially important for the star count method (see next section). We therefore average the positions and fluxes of the stars located $<2$\arcsec\ from each other.
We detect $\sim$ 70~000 distinct sources in $I$ and $J$ bands which are used to construct the extinction map (Section 3.1) and $\sim$ 20 000 sources detected in all 3 bands $IJK_s$.

\section{Selection}

\subsection{Extinction map of the Cha~II cloud}

We construct an extinction map of the Cha~II using the star count method developed by  \citet{cam99}. This method is based on the comparison of local stellar densities in the absorbed region and a nearby reference area assumed to be free of obscuration. DENIS provides stars detected in 2 infrared bands and 1 optical band. We first need to determine which band is the best one to apply this method. 
The $K_s$ band allows to probe the dense cores of the cloud but the density contrast is too small to build an extinction map.
We therefore use the $I$ and $J$ bands to construct the extinction map. By requiring that the stars be detected in both bands, we can eliminate spurious sources. But the simultaneous use of $I$ and $J$ bands introduces a bias which begins at $A_V$ = [$(I-J)_{\rm limit} - (I-J)_{\rm average}$]/[$<$$A_I/A_V$$>$$ - $$<$$A_J/A_V$$>$]. Fig. 2 shows that the average color of the stars $(I-J)_{\rm average}=1$. Using the values of $<$$A_I/A_V$$>$$ \rm and $$<$$A_J/A_V$$>$ from \citet{car89}, we find that the bias begins at $A_V$ = (2$-$1)/(0.479$-$0.282)=5. The bias can reach 2 magnitudes in the densest cores. As a result, less stars will be selected with a red color criterion.
 
We obtain $\sim$ 70 000 sources in $I$ and $J$ bands and use them to construct the extinction map. 
We use a wavelet transform algorithm to filter out the noise due the Poisson fluctuations in the star counts \citep[]{cam99}. We take into account the decrease of stellar density with the latitude to calibrate the extinction as a function of galactic latitude. The slope of $A_V(b)$ is equal to 0 as suggested by the morphologic similarity with the $^{13}CO$ map of \citet[]{miz99}.
The absolute calibration for the extinction is obtained by assuming the extinction is 0 outside of the cloud.
We use the extinction curve of \citet{car89} to convert $A_J$ into $A_V$ ($A_J= 0.282 \times A_V$). 

An alternative method to determine the extinction of individual stars consists of assuming an intrinsic color of stars and comparing this with the observed colors. However this method requires an assumption on the spectral types of the stars, of which we have no a priori information. The advantage of the star count method is that it provides an independent determination, at the expense of spatial resolution.

\begin{figure}
%\centerline{\psfig{file=contour.ps,width=8}
\psfig{file=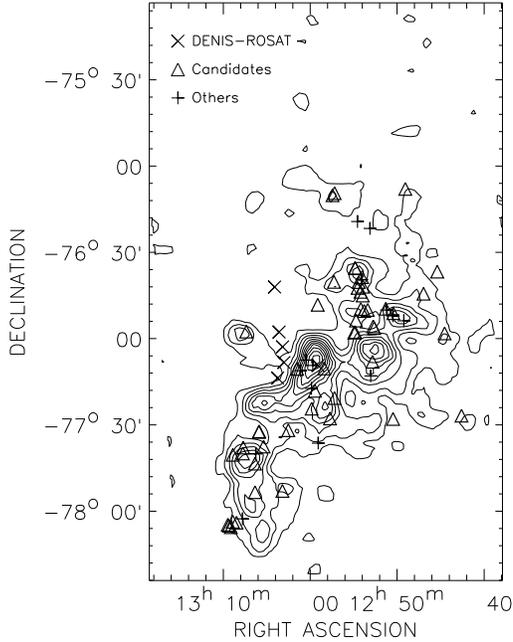,width=7cm}
\caption{ Spatial distribution of low-mass TTS and/or young brown dwarf candidates (triangle signs) selected with $I-J \ge 2$ in the Cha~II cloud. Six candidates cross-identified with ROSAT X-ray sources are presented. The plus signs are 15 bright candidates ($K_s \sim 8$). Contours indicate the extinction values beginning at $A_V=1$ and spaced by 1 as derived from DENIS $I$ and $J$ counts. Coordinates are J2000.}\label{extmap}
\end{figure}

Fig. \ref{extmap} shows the resulting extinction map of the Cha~II with a spatial resolution of 2$\arcmin$. Note that Fig. \ref{extmap} closely resembles the  $^{13}CO$ map of \citet[]{miz99}.
The extinction values A$_V$ range from 0 to 12. The statistical magnitude uncertainty due to the number of stars counted is approximately given by $1.3/\sqrt{(n+1)}$ \citep[]{ros80}, where $n$ is the number of stars counted per sampling element. Here, we have used $n=20$ and the statistical uncertainty is $\sim 0.30$ magnitudes in $A_J$ ($\sim 1.0$ in $A_V$). We use this map to deredden all stars detected by DENIS in 3 bands $IJK_s$ in this area.

\begin{figure}
%\vspace{-0.5cm}
\psfig{file=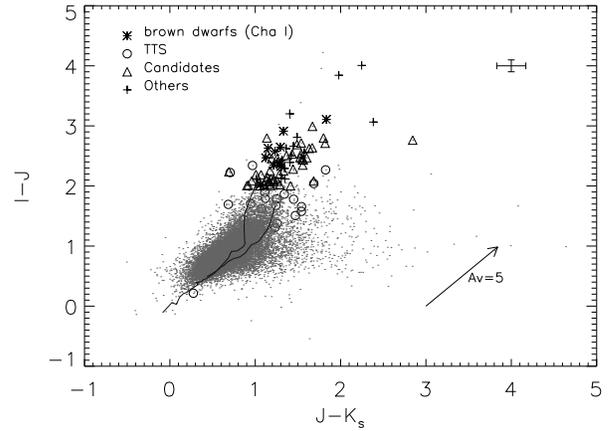,width=9cm}
\caption{$I-J/J-K_s$ color-color dereddened diagram for new DENIS selected low-mass TTS and/or young brown dwarf candidates (triangle signs) in Cha~II cloud. The dots represent 20~000 stars detected in DENIS $IJK_s$ bands in an area of 2.42$\degr$  $\times$ 3.50$\degr$ of the Cha~II. The open dot signs represent known TTS in the Cha~II. Eleven confirmed young brown dwarfs in Cha~I cloud are also shown (Comer\'on et al., 2000). The plus signs are 15 bright candidates ($K_s \sim 8$). The main sequence, the giant branch and the extinction vector are also plotted. The representative error bar is shown in the upper right corner.}\label{diag}
\end{figure}

\subsection{Selection criterion}

As shown by \citet{del97}, the $I-J$ versus $J-K_s$ diagram is a powerful tool to separate low-mass TTS and young brown dwarfs from main sequence stars. Brown dwarfs and low-mass TTS have no hydrogen fusion when they are young. Thus, the nature of these objects is the same, and only their mass and/or temperature are different. This explains why they are both characterised by a deficit in $I$ band because of the molecular absorption bands.  
Their youth is expressed by H${\alpha}$ emission and Li {\small I} absorption lines. But they are also characterised by features such as TiO and VO bands, and CaH, which allow the evaluation of the temperature and surface gravity. \citet{kir91} have shown that these bands are very sensitive to the temperature of the environment in the M dwarfs. Because these features are located in the $I$-band, these objects have a strong $I$ band deficit, while the $J$ band flux is essentially photospheric in all these objects. Therefore, $I-J$ is a good estimator of the effective temperature in these objects.   
Thanks to the association of the optical $I$ band to the infrared $J$ and $K_s$ bands in DENIS, we have derived a $I-J/J-K_s$ diagram (see Fig. \ref{diag}) for $\sim$20~000 sources detected in three bands $IJK_s$. We only select sources after dereddening with $I-J \ge 2$ and spatially distributed in groups around the cloud cores. This provides a sample of 98 sources. After visual inspection on the DENIS images, we exclude 28 objects because they are located near the bleed-out trails of nearby saturated stars.
We note that this selection criterion targets only young brown dwarfs and low-mass TTS. Massive TTS are characterised by their excess in $K_s$, thus in $J-K_s$ and $H-K_s$. Brown dwarfs and low-mass TTS have important $I-J$ but normal $J-K_s$. 

Table 1 lists the remaining 70 low-mass young star candidates with their intrinsic photometric uncertainties, ${\it i.e.}$ these do not include the uncertainties in the derived dereddening from the extinction map. Note that due to the bias (Sect. 3.1) introduced by $I$ and $J$ selection in the extinction map, less stars are selected in regions of high extinction. We identify 4 known TTS and 7 sources detected by IRAS within a 1$\arcmin$ radius in this sample. Four of our candidates are also detected by ISO (P. Persi, private communication). Fig. \ref{diag} shows these candidates, and 11 spectroscopically confirmed brown dwarfs detected in the Cha~I cloud \citep{com00}. We note that most of our candidates are located in the same region of the diagram as these known brown dwarfs.

\begin{figure*}
%\vspace{-0.3cm}
\psfig{file=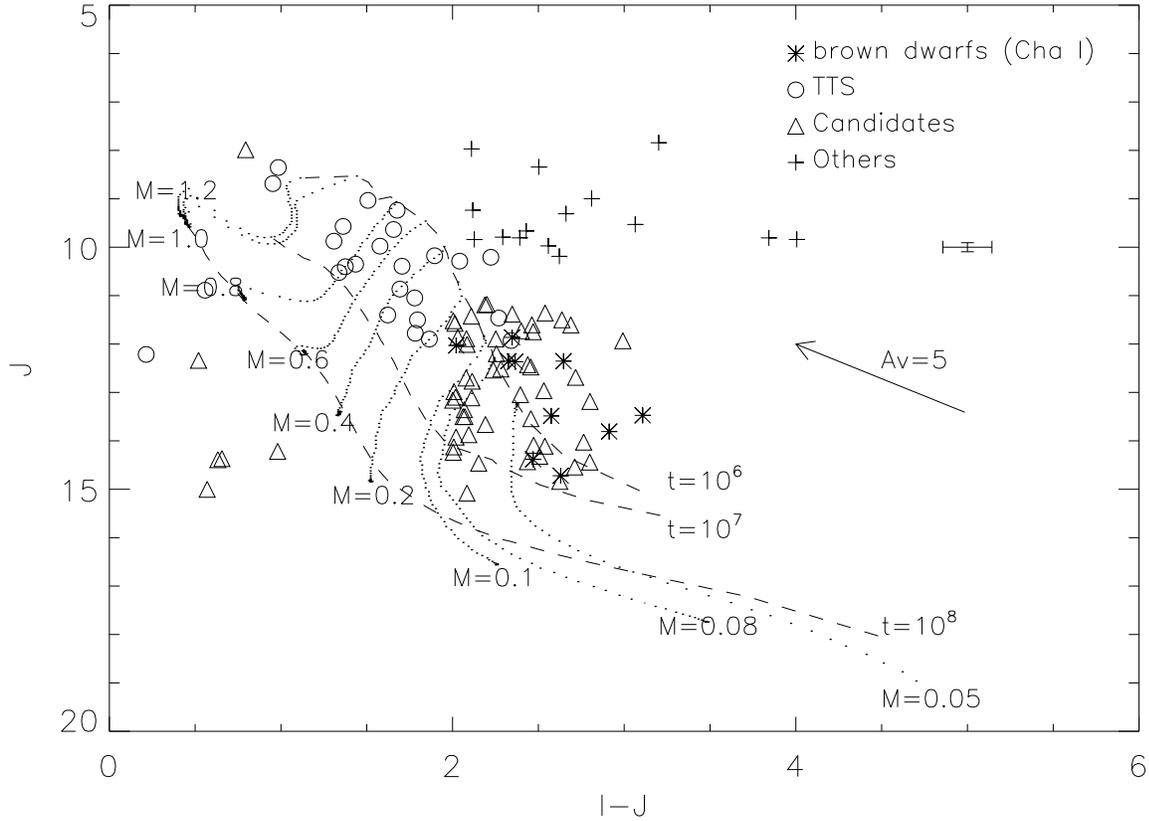,width=17cm}
%\centerline{\psfig{file=baraffe.ps}}
\caption{$J$ vs. $I-J$ magnitude-color diagram for the Cha~II dark cloud. Selected low-mass TTS and/or young brown dwarf candidates are indicated with triangles. We show the 1 Myr, 10 Myr and 100 Myr isochrones (dashed lines) from the \citet{bar98} model. Open circles denote known TTS of the Cha~II. Star symbols are brown dwarfs detected and spectroscopically confirmed by \citet{com00}. The plus signs are 15 bright candidates also  selected with $I-J \ge 2$. The representative error bar is given.}\label{bar}
%\vspace{-0.4cm}
\end{figure*}

\subsection{Cross-identification with ROSAT sources}

Weak line TTS are known as objects without a disk. 
Because of their relatively weak emission in the infrared (ascribed to the lack of dense surrounding matter), they can be more easily detected by their 
photospheric or coronal emission in X-ray surveys. 
Recently, \citet{alc00}, using ROSAT PSPC observations, have detected 40 X-ray sources in the Cha~II cloud, of which only 14 have been identified with previously known young stellar objects (IRAS sources, TTS). We cross-identified these 40 sources with DENIS $IJK_s$ data, and identified these 14 known objects, plus 6 additional new WTTS candidates. We have also checked that these sources are located in the cloud (see Fig. \ref{extmap}).

\section{Discussion and conclusion}

\subsection{Nature of the candidates}

From the comparison of these objects with known TTS and young brown dwarfs (Fig. \ref{bar}), we expect that most of our candidates are low-mass young stars such as young brown dwarfs or/and low-mass TTS. However, the true nature of the candidates is unambiguously confirmed only after a positive spectroscopic test such as the detection of the H$\alpha$ emission line or/and the Li {\small I} absorption line, signatures of the youth of the stars.
We will therefore observe these candidates spectroscopically. H$\alpha$ emission is not unique to young low-mass objects. The detection of Li~{\small I} from the photosphere can confirm the youth of the object \citep[]{mar99}. The detection of features such as CaH, TiI, NaI will allow the evaluation of the temperature and surface gravity \citep[]{kir91}. 

\subsection{Evolutionary status}

We use the evolutionary tracks for low-mass stars modeled by \citet{bar98} to estimate the mass and age of our low-mass star candidates. Fig.~\ref{bar} shows  the $J$ versus $I-J$ color-magnitude diagram of our 70 candidates selected by  $I-J \ge 2$ of which 4 are previously known TTS and six candidates cross-identified with ROSAT X-ray sources. We also plot the previously known TTS of the Cha~II cloud identified with DENIS sources and the brown dwarfs of the Cha~I detected and spectroscopically confirmed by \citet{com00}. We used our extinction map (Fig.~\ref{extmap}) to correct the reddening of all the objects shown in Fig.~\ref{bar}.
We include the isochrones at 1 Myr, 10 Myr and 100 Myr from the \citet{bar98} model. 
The selection criterion of sources with $I-J \ge 2$ does not allow to distinguish low-mass candidates and stars located just behind the cloud. Their location in the color-magnitude diagram shows that they are separated by their luminosity.
The location of low-mass candidates (except for the 15 bright candidates) in Fig.~\ref{bar} suggests that they, indeed, have masses $<$0.2~M$_{\sun}$, and ages between 1 and 10~Myr. Although we selected these sources only with the $I-J$ $\ge$ 2 criterion, they are remarkably close to the evolutionary tracks of the model. This confirms that they probably are young brown dwarfs.
Fifteen $I-J \ge 2$ candidates are bright sources ($K_s \sim$ 8). Their infrared color excess $J-K_s$ is $\sim$ $1.5-2$ and $I-J$ $\sim$ $2-4$. They do not appear to be field giants (expected to be uniformly distributed), but are located near the cloud cores in the extinction map. The number of giant field stars can be estimated using the so-called {\em Besan\c{c}on model} \citep[]{rob86}. For the total Cha~II area, the number of expected giant stars is about 400. A consequence of underestimating the extinction near the densest cores would be to select some giant stars near the cores. Only a spectroscopic follow-up can identify the nature of these objects.

\subsection{Conclusion}

We present the first extinction map of the Cha~II cloud using the DENIS $I$ and $J$ band. We use this map to deredden all stars detected in 3 bands $IJK_s$, which allows to select low-mass young stars embedded in the cloud cores.

The location of our dereddened candidates selected with $I-J \ge 2$ in the color-magnitude diagram suggests that 51 of our candidates are probably low-mass TTS and/or brown dwarfs. However, a spectroscopic follow-up is necessary to confirm the true nature of these objects.

This selection is independent from the characteristics of the dark cloud and can thus be applied to all other star formation regions observed by DENIS in the South Hemisphere. 

\begin{acknowledgements}
We are grateful to Dr. T. Montmerle for his helpful comments and to the anonymous referee for suggestions that improved this paper markedly.
We warmly thank the members of the DENIS consortium whose work made these results possible. The DENIS project is partly funded by the European Commission through {\it SCIENCE} and {\it Human Capital and Mobility} grants. It is also supported in France by INSU, the Education Ministry and CNRS, in Germany by the Land of Baden--W\"urtenberg, in Spain by DGICYT, in Italy by CNR, in Autria by the Fonds zur F\"orderung der Wissenschaftlichen Forschung and Bundesministerium f\"ur Wissenschaft und Forschung.
\end{acknowledgements}

%__________________________________________________ One column table
\begin{table*}
\caption[]{Photometric data of the selected candidates ($I-J \ge 2$)}
\small
\begin{tabular}{rrrrrrrrrr}
%\multicolumn{1}{c}{PAH} & \multicolumn{2}{c}{cation} &
%\multicolumn{2}{c}{neutral} \\
\hline
 ID$^a$  & $\alpha$(J2000) & $\delta$(J2000) & $I^b$ & $J^b$ & $K_s^b$ & $A_V$ & Ref.$^c$\\
\hline
 C1$\;\:$ & 12 43 53.1 &$-$77 29 44.6 &16.97$\pm$0.12 &14.67$\pm$0.12 &13.41$\pm$0.19 & 0.7 &\\
 C2$\;\:$ & 12 46 08.7 &$-$77 01 02.7 &17.77$\pm$0.44 &14.75$\pm$0.13 &13.43$\pm$0.17 & 1.1 &\\
 C3$\;\:$ & 12 47 11.7 &$-$76 39 29.8 &16.49$\pm$0.16 &14.34$\pm$0.11 &13.31$\pm$0.17 & 0.7 &\\
 C4$\;\:$ & 12 48 28.5 &$-$76 47 11.6 &17.31$\pm$0.20 &14.53$\pm$0.11 &12.74$\pm$0.12 & 1.5 &1\\
 C5$\;\:$ & 12 50 22.2 &$-$76 56 37.9 &16.51$\pm$0.08 &11.49$\pm$0.06 & 8.51$\pm$0.06 & 6.0 &\\
 C6$\;\:$ & 12 50 39.7 &$-$76 10 39.1 &17.32$\pm$0.12 &14.70$\pm$0.12 &12.98$\pm$0.14 & 1.0 &1\\
 C7$\;\:$ & 12 51 10.5 &$-$77 30 54.0 &17.26$\pm$0.03 &15.14$\pm$0.04 &13.42$\pm$0.08 & 0.2 &\\
 C8$\;\:$ & 12 51 28.3 &$-$76 54 00.3 &17.67$\pm$0.15 &14.56$\pm$0.12 &12.44$\pm$0.12 & 5.6 &5\\
 C9$\;\:$ & 12 51 29.0 &$-$76 54 55.5 &14.86$\pm$0.04 &11.29$\pm$0.06 & 9.01$\pm$0.06 & 5.8 &\\
C10$\;\:$ & 12 51 30.5 &$-$76 54 37.7 &14.42$\pm$0.04 &10.54$\pm$0.05 & 8.13$\pm$0.06 & 5.5 &\\
C11$\;\:$ & 12 52 11.0 &$-$76 52 53.3 &15.73$\pm$0.06 &10.95$\pm$0.05 & 8.04$\pm$0.07 & 3.9 &\\
C12$\;\:$ & 12 52 15.5 &$-$76 52 25.6 &17.51$\pm$0.14 &14.64$\pm$0.12 &13.05$\pm$0.16 & 3.5 &\\
C13$\;\:$ & 12 52 30.6 &$-$77 15 13.1 &12.21$\pm$0.02 & 9.10$\pm$0.05 & 7.25$\pm$0.08 & 4.1 &1,2\\
C14$\;\:$ & 12 53 23.4 &$-$76 59 11.3 &17.81$\pm$0.17 &14.90$\pm$0.14 &13.19$\pm$0.18 & 2.1 &\\
C15$\;\:$ & 12 53 29.7 &$-$76 58 34.9 &16.88$\pm$0.10 &14.41$\pm$0.12 &12.96$\pm$0.16 & 1.9 &\\
C16$\;\:$ & 12 53 29.8 &$-$77 10 56.8 &16.74$\pm$0.10 &13.43$\pm$0.09 &11.41$\pm$0.10 & 6.6 &5\\
C17$\;\:$ & 12 53 38.9 &$-$77 15 53.2 &15.25$\pm$0.05 &11.63$\pm$0.07 & 9.41$\pm$0.08 & 5.1 &1\\
C18$\;\:$ & 12 54 00.2 &$-$76 24 25.1 &11.52$\pm$0.02 & 8.12$\pm$0.05 & 6.55$\pm$0.09 & 1.0 &\\
C19$\;\:$ & 12 54 05.2 &$-$76 52 51.3 &16.13$\pm$0.07 &13.50$\pm$0.09 &11.88$\pm$0.11 & 2.6 &\\
C20$\;\:$ & 12 54 27.1 &$-$76 53 14.3 &16.86$\pm$0.11 &14.26$\pm$0.11 &12.65$\pm$0.14 & 2.7 &\\
C21$\;\:$ & 12 54 36.2 &$-$76 44 45.2 &17.99$\pm$0.18 &14.90$\pm$0.14 &13.28$\pm$0.19 & 2.8 &\\
C22$\;\:$ & 12 54 37.3 &$-$76 47 50.6 &16.37$\pm$0.09 &13.87$\pm$0.10 &12.40$\pm$0.11 & 2.5 &1\\
C23$\;\:$ & 12 54 42.6 &$-$76 41 25.8 &17.06$\pm$0.12 &13.73$\pm$0.09 &11.79$\pm$0.10 & 4.4 &\\
C24$\;\:$ & 12 54 45.2 &$-$76 42 19.5 &14.11$\pm$0.04 &10.99$\pm$0.05 & 8.93$\pm$0.07 & 4.2 &\\
C25$\;\:$ & 12 54 47.4 &$-$76 52 32.5 &16.89$\pm$0.11 &14.22$\pm$0.11 &12.54$\pm$0.12 & 3.0 &\\
C26$\;\:$ & 12 54 50.8 &$-$76 46 32.4 &16.80$\pm$0.10 &14.23$\pm$0.11 &12.74$\pm$0.14 & 2.6 &5\\
C27$\;\:$ & 12 55 03.3 &$-$76 43 07.8 &15.21$\pm$0.05 &12.26$\pm$0.07 &10.33$\pm$0.07 & 3.8 &\\
C28$\;\:$ & 12 55 08.7 &$-$76 45 10.4 &16.04$\pm$0.07 &13.44$\pm$0.08 &11.74$\pm$0.09 & 2.6 &\\
C29$\;\:$ & 12 55 14.4 &$-$76 22 00.5 &12.41$\pm$0.03 &10.10$\pm$0.05 & 8.64$\pm$0.06 & 0.9 &\\
C30$\;\:$ & 12 55 15.7 &$-$76 56 33.1 &15.45$\pm$0.06 &12.27$\pm$0.07 &10.14$\pm$0.07 & 2.7 &1\\
C31$\;\:$ & 12 55 20.7 &$-$77 00 35.9 &15.18$\pm$0.05 &12.11$\pm$0.07 &10.13$\pm$0.07 & 2.7 &\\
C32$\;\:$ & 12 55 24.0 &$-$76 38 17.6 &17.44$\pm$0.15 &13.88$\pm$0.10 &11.63$\pm$0.09 & 4.2 &\\
C33$\;\:$ & 12 55 25.7 &$-$77 00 46.5 &15.31$\pm$0.05 &12.58$\pm$0.07 &10.84$\pm$0.08 & 2.4 &\\
C34$\;\:$ & 12 55 25.7 &$-$76 38 10.9 &17.47$\pm$0.14 &14.24$\pm$0.11 &12.23$\pm$0.11 & 4.2 &\\
C35$\;\:$ & 12 57 25.9 &$-$77 23 35.2 &17.04$\pm$0.13 &14.18$\pm$0.10 &12.33$\pm$0.11 & 3.8 &\\
C36$\;\:$ & 12 57 29.3 &$-$76 12 06.5 &15.00$\pm$0.04 &12.49$\pm$0.07 &10.92$\pm$0.08 & 2.1 &\\
C37$\;\:$ & 12 57 31.5 &$-$76 43 04.4 &15.58$\pm$0.06 &12.55$\pm$0.07 &10.47$\pm$0.07 & 2.9 &1\\
C38$\;\:$ & 12 57 41.6 &$-$76 12 51.2 &15.96$\pm$0.07 &13.07$\pm$0.08 &11.13$\pm$0.08 & 2.3 &\\
C39$\;\:$ & 12 57 54.5 &$-$77 30 43.5 &15.75$\pm$0.06 &12.73$\pm$0.07 &10.73$\pm$0.07 & 4.6 &\\
\end{tabular}
\end{table*}

\begin{table*}
\small
\begin{tabular}{rrrrrrrrrr}
%\multicolumn{1}{c}{PAH} & \multicolumn{2}{c}{cation} &
%\multicolumn{2}{c}{neutral} \\
\hline
 ID$^a$  & $\alpha$(J2000) & $\delta$(J2000) & $I^b$ & $J^b$ & $K_s^b$ & $A_V$ & Ref.$^c$\\
\hline
C40$\;\:$ & 12 58 31.4 &$-$77 13 16.4 &17.73$\pm$0.17 &14.00$\pm$0.10 &11.49$\pm$0.08 & 8.8 &\\
C41$\;\:$ & 12 59 09.8 &$-$76 51 03.5 &16.82$\pm$0.10 &14.04$\pm$0.10 &11.19$\pm$0.09 & 0.1 &1\\
C42$\;\:$ & 12 59 10.7 &$-$77 39 19.1 &12.91$\pm$0.02 & 9.86$\pm$0.05 & 8.08$\pm$0.06 & 2.0 &\\
C43$\;\:$ & 12 59 30.3 &$-$77 21 04.6 &17.27$\pm$0.12 &13.97$\pm$0.09 &11.66$\pm$0.09 & 5.2 &\\
C44*      & 12 59 42.8 &$-$77 12 12.8 &15.68$\pm$0.06 &11.27$\pm$0.05 & 8.29$\pm$0.05 &11.7 &\\
C45$\;\:$ & 12 59 51.8 &$-$77 27 18.9 &16.20$\pm$0.08 &12.86$\pm$0.07 &10.56$\pm$0.07 & 4.5 &\\
C46*      & 12 59 52.1 &$-$77 20 15.0 &14.30$\pm$0.03 &10.97$\pm$0.06 & 8.59$\pm$0.06 & 6.2 &\\
C47*      & 13 00 24.1 &$-$77 10 22.4 &17.18$\pm$0.12 &12.71$\pm$0.07 & 9.51$\pm$0.06 & 9.7 &\\
C48*      & 13 00 59.3 &$-$77 14 02.7 &16.40$\pm$0.08 &11.77$\pm$0.09 & 8.05$\pm$0.08 & 7.9 &\\
C49*      & 13 01 21.4 &$-$77 13 25.2 &17.62$\pm$0.15 &13.52$\pm$0.11 &10.90$\pm$0.10 & 5.6 &\\
C50$\;\:$ & 13 02 22.9 &$-$77 34 49.5 &15.42$\pm$0.05 &12.77$\pm$0.10 &11.26$\pm$0.11 & 2.0 &\\
C51$\;\:$ & 13 03 09.1 &$-$77 55 59.5 &14.46$\pm$0.04 &11.83$\pm$0.07 &10.38$\pm$0.08 & 2.3 &\\
C52$\;\:$ & 13 05 05.1 &$-$77 40 31.3 &17.34$\pm$0.14 &13.98$\pm$0.10 &11.71$\pm$0.10 & 2.8 &\\
C53$\;\:$ & 13 05 08.6 &$-$77 33 42.5 &14.68$\pm$0.04 &12.02$\pm$0.06 & 9.87$\pm$0.07 & 2.0 &3\\
C54$\;\:$ & 13 05 30.9 &$-$77 35 11.9 &17.10$\pm$0.11 &14.75$\pm$0.12 &13.03$\pm$0.17 & 1.8 &\\
C55$\;\:$ & 13 05 32.7 &$-$77 35 26.2 &14.60$\pm$0.04 &11.89$\pm$0.06 &10.03$\pm$0.06 & 1.8 &\\
C56$\;\:$ & 13 05 57.1 &$-$77 43 00.5 &13.74$\pm$0.03 &10.05$\pm$0.04 & 7.61$\pm$0.05 & 6.0 &\\
C57$\;\:$ & 13 06 04.8 &$-$77 46 28.3 &16.75$\pm$0.09 &13.54$\pm$0.08 &11.38$\pm$0.08 & 6.0 &\\
C58$\;\:$ & 13 06 11.1 &$-$77 56 26.4 &17.64$\pm$0.15 &14.92$\pm$0.13 &13.04$\pm$0.17 & 3.5 &\\
C59$\;\:$ & 13 06 30.8 &$-$77 00 24.0 &15.96$\pm$0.07 &12.59$\pm$0.07 &10.52$\pm$0.07 & 3.5 &\\
C60$\;\:$ & 13 06 56.5 &$-$77 23 09.5 &13.59$\pm$0.03 &10.89$\pm$0.05 & 9.78$\pm$0.06 & 2.4 &4\\
C61$\;\:$ & 13 06 57.4 &$-$77 23 41.5 &13.56$\pm$0.03 &11.01$\pm$0.05 & 8.90$\pm$0.06 & 2.6 &1,4\\
C62$\;\:$ & 13 07 18.1 &$-$77 40 53.0 &16.07$\pm$0.07 &13.17$\pm$0.07 &11.52$\pm$0.08 & 4.1 &\\
C63$\;\:$ & 13 07 21.4 &$-$77 42 55.6 &17.81$\pm$0.20 &14.69$\pm$0.12 &12.64$\pm$0.12 & 5.6 &5\\
C64$\;\:$ & 13 07 41.2 &$-$78 05 43.3 &13.17$\pm$0.03 &10.38$\pm$0.05 & 8.63$\pm$0.06 & 2.0 &\\
C65$\;\:$ & 13 08 22.8 &$-$78 07 04.6 &14.95$\pm$0.04 &12.21$\pm$0.06 &10.66$\pm$0.07 & 1.7 &\\
C66$\;\:$ & 13 08 27.2 &$-$77 43 23.3 &16.53$\pm$0.09 &13.57$\pm$0.08 &12.26$\pm$0.10 & 3.7 &\\
C67$\;\:$ & 13 08 46.8 &$-$78 06 44.5 &16.50$\pm$0.09 &13.84$\pm$0.09 &12.40$\pm$0.12 & 1.0 &1\\
C68$\;\:$ & 13 09 02.4 &$-$78 08 38.6 &15.95$\pm$0.07 &13.23$\pm$0.07 &11.64$\pm$0.08 & 0.9 &\\
C69$\;\:$ & 13 09 14.6 &$-$78 08 06.3 &17.62$\pm$0.18 &14.75$\pm$0.12 &12.81$\pm$0.15 & 0.8 &\\
C70$\;\:$ & 13 09 21.6 &$-$78 07 47.8 &17.81$\pm$0.20 &15.04$\pm$0.14 &13.29$\pm$0.19 & 0.7 &\\
\hline
 X1$\;\:$ & 12 59 10.1 &$-$77 12 13.7 &14.29$\pm$0.03 &11.23$\pm$0.06 & 9.19$\pm$0.06 &11.5 \\
 X2$\;\:$ & 13 02 42.3 &$-$77 11 11.1 &15.68$\pm$0.03 &14.75$\pm$0.09 &13.50$\pm$0.13 & 1.4 \\
 X3$\;\:$ & 13 02 49.2 &$-$77 05 44.6 &15.31$\pm$0.05 &14.28$\pm$0.13 &13.33$\pm$0.19 & 0.2 \\
 X4$\;\:$ & 13 03 09.0 &$-$77 00 32.8 &13.18$\pm$0.03 &12.53$\pm$0.08 &11.96$\pm$0.11 & 0.7 \\
 X5$\;\:$ & 13 03 23.9 &$-$77 16 37.1 &15.45$\pm$0.05 &14.64$\pm$0.12 &13.46$\pm$0.22 & 0.9 \\
 X6$\;\:$ & 13 03 30.5 &$-$76 44 51.4 &15.57$\pm$0.05 &15.00$\pm$0.14 &13.50$\pm$0.22 & 0.0 \\
\hline
\end{tabular}
\\
$^a$ C=Colour selected candidate; X=X-ray selected candidate; *=object\\ detected by ISO (P. Persi, private communication)\\
$^b$ Observed magnitudes. The intrinsic photometric uncertainty is given for each star.\\
$^c$ REFERENCES: (1): IRAS; (2):\citet{lar98}; (3):\citet{har93}; (4):\citet{sch77}; (5): \citet{vil94}
%\vspace{-0.6cm}
\end{table*}
\normalsize

\end{document}